\documentclass{ws-procs9x6}

\usepackage{graphicx}

\begin{document}

\title{Double Longitudinal Spin Asymmetry in Inclusive Jet Production in Polarized p+p Collisions at $\sqrt{s}$ = 200 GeV }

\author{F. Simon (for the STAR Collaboration)}

\address{Massachusetts Institute of Technology \\
77 Massachusetts Avenue \\ 
Cambridge, MA 02139, USA\\ 
E-mail: fsimon@mit.edu}
\maketitle

\abstracts{
We present preliminary measurements of the cross section and the double longitudinal spin asymmetry in inclusive jet production in polarized p+p collisions at $\sqrt{s}$ = 200 GeV. The measured cross section agrees well with NLO pQCD calculations over seven orders of magnitude. The observed spin asymmetries are consistent with theoretical evaluations based on deeply inelastic scattering data and tend to disfavor a large positive gluon polarization.}

\section{Introduction}
The Relativistic Heavy Ion Collider RHIC is the first polarized high-energy proton-proton collider, providing polarized p+p collisions at energies up to $\sqrt{s}$ = 500 GeV. One of the main objectives of the RHIC-SPIN program is the precise determination of the gluon contribution to the spin of the nucleon via the measurement of double longitudinal spin asymmetries $A_{LL} = \Delta\sigma / \bar{\sigma} = (\sigma^{++} - \sigma^{+-})/(\sigma^{++} + \sigma^{+-})$ in a variety of processes covering a wide kinematic range of $0.01 < x < 0.3$\cite{SpinRev}. The processes under study encompass inclusive jet and pion production, di-jets and di-hadrons, production of heavy quark pairs and photon-jet coincidences.  One of the measurements with only modest requirements on the total integrated luminosity is the measurement of the spin asymmetries of inclusive jet production. 

With its large acceptance tracking and electromagnetic calorimetry the STAR detector\cite{STAR} is uniquely capable of full jet reconstruction in p+p collisions at RHIC. The dataset discussed here was recorded during commissioning runs in 2003 and 2004, aimed at developing luminosity and polarization. The recorded luminosity in those two runs is $\sim$0.5 pb$^{-1}$, with average polarizations of 30\% -- 40\%, allowing an exploratory study of spin asymmetries in inclusive jet production. A first measurement of the inclusive jet cross section is obtained using the $\sim$0.2 pb$^{-1}$ recorded in 2004.

\section{Jet Analysis and Inclusive Cross Section}

The STAR detector subsystems of principal interest for this analysis are the plastic scintillator beam-beam Counters (BBC), the time projection chamber (TPC) and the barrel electromagnetic callorimeter (BEMC). The BBCs cover $3.3<\vert\eta\vert<5.0$ in pseudorapidity and are used to trigger on non-singly diffractive (NSD) inelastic reactions. This minimum bias (MB) trigger accepts $\sim$87\% of the NSD cross section, corresponding to $26.1\pm 2.0$ mb\cite{BBCXSect}. The TPC provides charged particle tracking over $\vert\eta\vert<1.2$ in full azimuth. The BEMC, partially commissioned in 2004 (2400 out of 4800 towers, full azimuthal coverage) covered $0<\eta<1$. In addition to the MB trigger, a high tower (HT) trigger that requires one calorimeter tower above an $E_T$ threshold corresponding to $\sim$2.5 ($\sim$3.0) GeV at $\eta$ = 0 (1) on top of the MB condition was used. This trigger detects energetic $\pi^0$, $\gamma$ and electrons. In order to enrich the sample of highly energetic processes, the MB trigger was highly prescaled during data taking.

Events were accepted for analysis if the primary vertex was within 60 cm of the center of the detector along the beam axis in order to ensure a uniform tracking efficiency. For HT triggered events an $E_T$ $>$ 3.5 GeV in the trigger tower was required to ensure uniform trigger efficiency over the full BEMC acceptance. Jet finding was performed with the midpoint-cone algorithm\cite{Midpoint}, using all charged tracks originating from within 3 cm of the primary vertex and all calorimeter towers. To reject background events, a sizable contribution from charged tracks to the total jet energy was required. To avoid edge effects at the extreme ends of the detector acceptance, the jet axis was required to be within $0.2 < \eta < 0.8$. 

\begin{figure}[t]
\begin{center}
\includegraphics[width=0.65\linewidth]{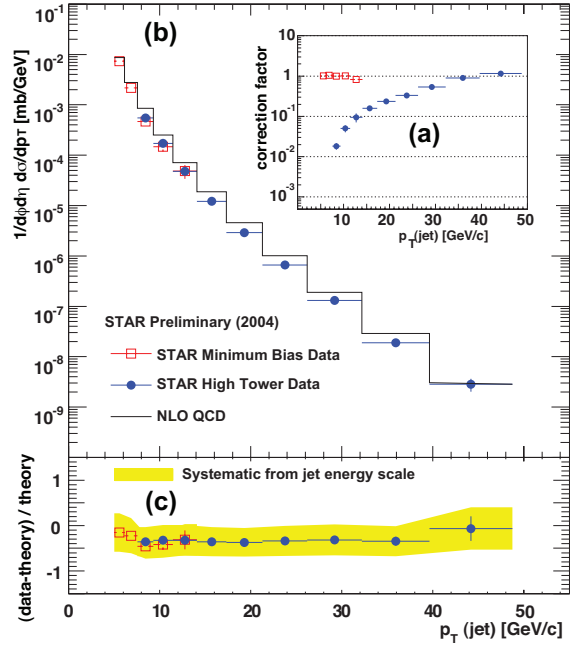} 
\end{center}
\caption{(a) The correction factor for MB and HT data obtained from PYTHIA simulations. Statistical
errors are shown. (b) Preliminary inclusive jet cross section compared to NLO pQCD calculation. Statistical uncertainties only. (c) Ratio comparison of data vs. theory. The shaded band represents the dominant systematic uncertainty from the jet energy scale, and an 8\% overall normalization uncertainty is not shown. See text for details. 
\label{fig:XSect}}
\end{figure}

In order to extract the inclusive jet cross section a correction for detector efficiency and resolution is required. The bin-by-bin correction factors were obtained by studying PYTHIA (v6.205)\cite{Pythia} events passed through a GEANT model, a response simulator of the STAR detector and the full data reconstruction frame work. The jet $p_T$ resolution was determined to be 25\%, which motivated the choice of the binning. A separate determination of the correction factors was done for MB and HT events. While the MB correction factor is close to unity, the HT correction factor is dominated by the trigger efficiency and varies over two orders of magnitude with $p_T$. The correction factors for both triggers are shown in figure \ref{fig:XSect}a. Figure \ref{fig:XSect}b shows the corrected cross section. In the three overlapping bins, good agreement is seen between MB and HT triggered events. The data are compared to NLO pQCD calculations incorporating CTEQ06M PDFs with $\mu_F = \mu_R = p_T$\cite{GRSV_Jet}. Figure \ref{fig:XSect}c shows the ratio of data - theory divided by theory. The shaded band indicates the dominant systematic uncertainty (50\% change in yield) introduced by a 10\% uncertainty in the jet energy scale. For $p_T$ $>$ 10 GeV there is a systematic offset between data and theory, but there is good agreement within the large systematic uncertainties. It is also apparent that the spectral shape of the data agrees very well with the theory predictions. Improved calibrations and future measurements of di-jets and $\gamma$-jet coincidences are expected to reduce the uncertainties of the jet energy scale.

\section{Spin Asymmetry}

The cross section for inclusive jet production depends on the spin configuration of the two colliding protons. In leading order, $gg \rightarrow gg$, $gq \rightarrow gq$ and $qq \rightarrow qq$ subprocesses contribute to the cross section. The relative contribution of these subprocesses to the spin asymmetry in the cross section is $p_T$ dependent. The dominant contributions in the $p_T$ region covered by STAR are from $gg$ and $qg$ scattering, with $gg$ subprocesses dominating at low $p_T$ and $qg$ at higher $p_T$\cite{GRSV_Jet}.

The double longitudinal spin asymmetry is defined as 
\begin{equation}
A_{LL}^{jet} = \frac{1}{P_1P_2}\,\frac{N^{++} - RN^{+-}}{N^{++} + RN^{+-}}
\end{equation}
where $N^{++(+-)}$ are the inclusive jet yields for equal (opposite) spin orientations of the colliding protons, R = $L^{++}/L^{+-}$ is the ratio of luminosities for the different spin configurations and $P_{1(2)}$ are the polarizations of the two proton beams. The relative luminosities and the orientation of the polarization vector in the STAR interaction region is measured with the BBCs. The beam polarization is obtained with the RHIC polarimeters\cite{RhicCNI}.  

\begin{figure}[t]
\begin{center}
\includegraphics[width = 0.7\linewidth]{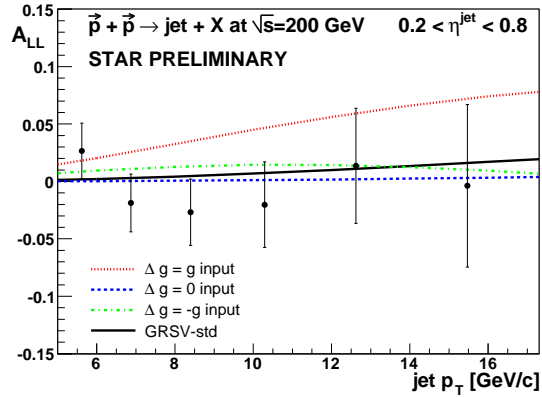}
\end{center}
\caption{Double longitudinal spin asymmetry $A_{LL}$ for inclusive jet production in polarized p+p collisions at 200 GeV as a function of jet $p_t$. Errors are statistical only, systematic uncertainties (without polarization) are $\sim$0.01. The curves show predicted asymmetries for different gluon polarizations, see text for details. \label{fig:All}}
\end{figure}

Figure \ref{fig:All} shows preliminary results for the double longitudinal spin asymmetry $A_{LL}$ in inclusive jet production for the datasets taken in 2003 and 2004. Systematic uncertainties originating from the relative luminosity, trigger bias, beam background and non-longitudinal spin contributions have been investigated. Studies of parity violating single spin asymmetries and randomized spin patterns show no evidence for bunch to bunch or fill to fill systematics. The curves in figure \ref{fig:All} show theoretical evaluations of $A_{LL}^{jet}$ in inclusive jet production for different sets of polarized gluon distributions functions\cite{GRSV_Jet,GRSV}. The GRSV-std curve is based on the best fit to DIS data, the other curves use a vanishing gluon polarization $\Delta g(x, Q^2_0) = 0$ and maximally positive and negative gluon polarization $\Delta g(x, Q^2_0) = \pm g(x, Q^2_0)$ at an input scale of $Q^2_0$ = 0.6 GeV$^2$/c$^2$. The data are consistent with three of these evaluations and tend to disfavor the scenario with a large positive gluon polarization.

So far, the measurements are limited by statistical precision. The data taken in 2005 are in the final stages of analysis and will provide a significant reduction of statistical errors due to higher polarization and an order of magnitude increase in sampled luminosity. Further improvement is expected with increasing integrated luminosity and improved polarization in future runs. 

\section{Conclusion}

We have presented preliminary measurements of the cross section and the double longitudinal spin asymmetry in inclusive jet production in polarized p+p collisions at $\sqrt{s}$ = 200 GeV. The cross section is in reasonable agreement with NLO pQCD calculations over seven orders of magnitude, motivating the application of these calculations to interpret the measured spin asymmetries. The preliminary double-longitudinal spin asymmetry is consistent with an evaluation based on a fit to DIS results and disfavors large positive values for the gluon polarization.

\end{document}